\documentstyle[12pt]{article}
\catcode`\@=11
\def\vereq#1#2{\lower3pt\vbox{\baselineskip1.5pt \lineskip1.5pt
\ialign{$\m@th#1\hfill##\hfil$\crcr#2\crcr\sim\crcr}}}

\catcode`\@=12
\begin{document}

\begin{center}
{\bf COLOR TRANSPARENCY PHYSICS: A NEW DOMAIN OF STRONG INTERACTIONS
\footnote{ to
be published in the Proceedings of the Workshop on Quantum Chromodynamics,
Paris June
1996, World Scientific ed.}}
\\
\vspace{10mm}

 Bernard PIRE
\\
\vspace{4mm}

 {\it Centre de Physique Th\'eorique {\footnote {unit\'e propre 014 du
CNRS}}, Ecole Polytechnique
\\ F91128 Palaiseau, France}
\\
\vspace{8mm}
ABSTRACT
\end{center}

{\leftskip=12mm \rightskip=12mm \footnotesize \baselineskip=12pt
\noindent
The concept of color transparency is introduced. This new feature of QCD is
characteristic of a gauge theory. It enables strong interactions to be studied
in a new domain: scattering amplitudes of transversally small color singlet
objects. Experimental data are still few and somehow
inconclusive. Future prospects are briefly discussed. This is a frontier
problem of QCD which deserves much attention from both theorists and
 experimentalists in the coming years.
\par}

\vspace{10mm}

The current lively debate on the nature of the Pomeron reflects our poor
understanding
of long distance strong interaction physics. Color transparency experiments
offer us in
this respect a unique opportunity to challenge various theoretical models
through their
 description of the strong interactions of color singlet {\em small} objects.

\section{ The idea }
\hspace {\parindent}
The concept of color  transparency~\cite{CT} has recently attracted much
attention.
This phenomenon illustrates the power of exclusive reactions to  isolate
simple elementary quark configurations. The experimental technique  to
probe these configurations is the following.

 For  a hard  exclusive reaction,  say electron  scattering from  a
proton, the  scattering amplitude  at large  momentum transfer  $Q^2$ is
suppressed  by  powers  of  $Q^2$  if  the proton contains more than the
minimal number  of constituents.   This  is derived  from the  QCD based
quark  counting   rules,  which   result  from   the  factorization   of
wave-function-like  distribution  amplitudes.    Thus protons containing
only  valence  quarks  participate  in  the  scattering.  Moreover, each
quark,  connected  to  another  one  by  a  hard gluon exchange carrying
momentum of order $Q$, should be found within a distance of order $1/Q$.
Thus, at  large $Q^2$ one  selects a very  special quark configuration which
can be defined in terms of selected short distance regions  of the wave
function:
all connected  quarks are  close together,  forming a  small size  color
neutral  configuration. While this is just an important region, quantum
mechanically contributing to the process, it is  sometimes  referred  to
 as a {\em mini hadron}. We will use this term freely, with the
understanding that
the term refers to a region, not a hadron.
This mini hadron (like any amplitude selected for its importance)
is not a  stationary state and  evolves until finally one measures
combinations of
normal hadrons.

Such  a color  singlet system  cannot emit  or absorb  soft gluons
which carry energy or momentum smaller than $Q$.  This is because  gluon
radiation --- like photon radiation in QED --- is a coherent process and
there is thus destructive interference between gluon emission amplitudes
by quarks  with ''opposite''  color.   Even without  knowing exactly how
exchanges  of   soft  gluons   and  other   constituents  create  strong
interactions, we  know that  these interactions  must be  turned off for
small color singlet objects.

An exclusive hard reaction will thus probe strong interactions in three
complementary ways:

\begin{itemize}

\item  First, selecting the simplest Fock state amounts to the study of
 the structure of a {\em  mini hadron}, i.e. the short distance part of a
 minimal Fock state  component in the hadron  wave function. This is of
 primordial interest for  the understanding  of  the  difficult  physics
 of  confinement.

\item  Secondly,
letting the mini-state evolve during its travel through different nuclei
of various  sizes allows  an indirect  but unique  way to  test how  the
squeezed mini-state  goes back  to its  full size  and complexity,  {\em
i.e.} how  quarks inside  a color-singlet state  rearrange themselves
spatially to
''reconstruct'' a normal size hadron.   In this respect the  observation
of baryonic resonance  production as well  as detailed spin  studies are
mandatory.

\item Thirdly, the study of the (reduced) final state interactions of the
 {\em  mini hadron} with spectator hadrons, through the soft scattering of these
 latter ones, opens the domain of the strong interactions of small color-singlet
objects, much related to Pomeron physics.
\end{itemize}

\section {\bf Present Data}

\hspace {\parindent}
Experimental data on color transparency are very scarce but worth
considering in detail. The first piece of evidence for something like
color transparency came from the Brookhaven experiment on pp elastic
scattering at $90^\circ$  CM in a nuclear medium ~\cite{ref:ASC88}{}.
These data
lead to a lively debate. The special feature of hadron hadron elastic scattering
at fixed angle is that in addition to a clear cut  short distance amplitude,
there is an infrared sensitive process (the so-called
independent scattering mechanism) which allows not so small configurations to
scatter elastically. The phenomenon of colour transparency is thus
replaced by a {\it nuclear filtering} process: elastic scattering in
a nucleus filters away the big component of the nucleon wave function
and thus its contribution to the cross-section. Since the presence of these
two competing processes had been analysed \cite{ref:PR82}{} as responsible for
the oscillating pattern seen in the scaled cross-section $s^{10}d\sigma/dt$,
an oscillating color transparency ratio emerges~\cite{ref:RP90}.
One way ("attenuation method") to understand data is to define a survival
probability
 related in a standard way to an effective attenuation cross section
 $\sigma_{eff}(Q^2) $ and to plot this attenuation cross section as a
 function of the transfer of the reaction\cite{ref:JR}{}. One indeed obtains
 values of  $\sigma_{eff}(Q^2) $ decreasing with $Q^2 $ and quite smaller
 than the free space inelastic proton cross section. The survival probability
is even found to obey a simple scaling law in $Q^2/A^{1/3}$\cite{ref:PR}{}

The SLAC NE18 experiment\cite{ref:SLAC}{} recently
 measured the color transparency ratio up to
$Q^2=7 GeV^2$ , without any observable increase. This conclusion follows
only if
the hard scattering part of the process is assumed to be the same as in
free space,
canceling out in forming the transparency ratio. While this assumption is
not needed
in the attenuation method, the precision of the data were not sufficient to
conclude much using the less model-dependent test. While the majority view
is that
these data cast doubt on the most optimistic onset of color transparency,
emphasizing
 the importance of a sufficient boost to get the small state quickly out of
the nucleus,
this conclusion remains tentative and something to be tested.

The diffractive electroproduction of vector mesons at
 Fermilab~\cite{ref:Fermilab}{} and Cern~\cite{ref:NMC}{}  exhibit
an interesting increase of the transparency ratio for data at $Q^2 \simeq 7
GeV^2$.
In this case the boost is high since the lepton energy is around
$E \simeq 200 GeV$ but the problem is to disentangle diffractive
from inelastic events.

\section{\bf Future prospects }
\hspace {\parindent}
It should by now be obvious to the reader that Color Transparency is just an
emerging field of study and that one should devote much attention to get
more information on this physics in the near future.
\subsection{\bf Experiments }
A second round of proton experiments at Brookhaven has been approved and
 a new detector named EVA ~\cite{BNL850} with much higher acceptance
 has been taking
data for about one year. Along with other improvements and increased
beam type, this should increase the amount of data taken by a factor
 of 400 allowing a larger energy range and an analysis at different
scattering angles. It may also open the field of  meson-nucleus reactions.

The Hermes detector ~\cite{HERMES} at HERA is beginning operation. It will
enable a
confirmation of existing data on $\rho$ meson diffractive production at
moderate
$Q^2$ values and quite smaller values of energies:
 $10 \leq \nu \leq 22 GeV$.

\noindent
This experiment might however suffer from the same weakness as
the ones from FNAL and CERN since Hermes small luminosity only allows integrated
measurements and thus cannot assure that  diffractive events are not
polluted by inelastic events. It seems difficult to envisage in the
near future the detection of the recoiling proton.

The $15-30 GeV$ continuous electron beam ELFE project is presently discussed
at the European level~\cite{ref:ELFE}. Besides the determination of
hadronic valence
wave functions through the careful study of many exclusive hard reactions
in free space, the use of nuclear targets to test and use  color
transparency is one of
 its major goals. The $(e,e',p)$ reaction should in particular be studied
in a wide
range of $Q^2$ up to $21 GeV^2$,thus allowing to connect to SLAC data (and
better
 resolution but similar low  $Q^2$ data from CEBAF) and hopefully clearly
establish
 this phenomenon in the simplest occurence.
The measurement of the transparency ratio for photo-and
electroproduction of heavy vector mesons, in particular of $\psi$ and
 $\psi'$ will open  a new regime where the mass of the quark enters as
an other scale controlling the formation length of the produced meson.
\subsection{\bf Theory }

\subsubsection{Large {\it vs} Small Nuclei }

To the extent that the electromagnetic form factors are understood as  a
function of $Q^2$,
 $eA \rightarrow e' (A-1)  ~ p$
experiments will measure
the color screening properties of QCD.   The most straightforward quantity to be
 measured  is the transparency ratio $T_r$ which is defined as:
\begin{equation}
T_r = \frac{\sigma_{Nucleus}}{Z \sigma_{Nucleon}}
\end{equation}

At asymptotically large values  of $Q^2$, dimensional estimates  suggest
that $T_r$ scales as a function of $A^{\frac{1}{3}}/Q^2$~\cite{ref:PR}{}.
The  approach to the scaling behavior as well as  the value of $T_r$ as a
function  of
the  scaling  variable  determine  the  evolution  from  the  mini-region
 to the  complete hadron.   This interesting  effect
can be measured in an $(e , e' p)$ reaction that  provides
the best chance for a {\it quantitative} interpretation.

Although color transparency was first mostly considered as an effect to be
studied  on rather large nuclei, it became  recently clear that small nuclei
had much to teach us about this physics item. Deuteron  electrodesintegration
reactions $d(e,e'p)n$ for instance~\cite{deut}, both polarized and unpolarized,
is much interesting. The idea is  simple; let us examine the case where the
 virtual photon mostly hit the proton in the deuteron. The
   neutron momentum distribution is then  due to the combination of
\begin{itemize}
\item  Fermi momentum effects in the initial state;
\item  final state interactions.
\end {itemize}
\noindent
These two components are well known at small $Q^2$.
At large  $Q^2$, the hard process selects small-sized proton, and the
interaction {\it miniproton}-neutron is much weaker. This is where Pomeron
 physics enters, at least if energy is high enough. High $Q^2$
electroproduction
data appear then as an unexpected testing bench of soft physics, with the
important bonus of a controlable variable sized hadron scattering on a
normal one.
Whether the small size justifies completely a  perturbative treatment of
this small transfer amplitude is still an open question.

The cross section for  $d(e,e'p)n$ may be written as

\begin{equation}
{d\sigma \over {dE_{e'}d\Omega_{e'}d^3p_p }} = \sigma_{ep} ~~D_d(q,p_p,p_n) ~~
\delta (q_0-M_d-E_p-E_n)
\end{equation}

\noindent
where $D_d(q,p_p,p_n)$ is the joint probability of the
 initial proton having a Fermi momentum $p_p-q$ and the final proton (neutron)
a momentum $p_p$ ($p_n$). A poor energy resolution would restrict the physics
to a qualitative
 observation of color transparency, whereas a very good one would allow a
quantitative determination of the  miniproton-neutron scattering cross-section
as a  function of the miniproton size ({\it i.e.} $Q^2$), provided one
controls and
deconvoluates Fermi momentum effects in the deuteron.

Frankfurt {\it et al}~\cite{deut} estimate sizable  effects at CEBAF
energies, which amount to
$Q^2$ values in the range
 $\sim 4 GeV^2 \leq Q^2\leq~10~(GeV/c)^2$. Prospects are brighter within
 ELFE conditions.

\subsubsection{Helicity (non-)conservation }
 The hadronic helicity conservation rule in hard exclusive reactions~\cite{hel}
 follows from two assumptions:
\begin{itemize}
\item quark masses can be  neglected;
\item  valence states (with only fermions) dominate.

\end {itemize}
\noindent
These assumptions which asymptotically are quite solid in reactions such as
 $(e,e',p)$ are less justified in the hadronic case~\cite{gpr}. They are
exactly
what leads to the result of color transparency. It thus follows that
{\it the helicity non-conserving contributions must be  filtered away
 in a nuclear medium}.
It is thus most interesting, at a given value of  $Q^2$ to compare
the nuclear absorbtion of  amplitudes  violating the helicity conservation rule.
 For this measurement to be possible, we must consider cases where such
amplitudes
are quite large at reasonable $Q^2$ values. This is not the case
for the proton form factor $F_2$, but maybe for the $ p-\Delta$ transition form
factor~\cite {Burkert}, measured at $Q^2 = 3.2 GeV^2$.

In the case of hadronic reactions, it has been predicted~\cite{gpr} that
the amount
of helicity non conservation seen for instance in the helicity matrix
elements of the  $\rho$ meson produced in
$\pi p\rightarrow\rho p$ at 90$^\circ$ would be filtered out in a nucleus.
 Experimental data in free space~\cite{hep} yield
$\rho_{1-1}=0.32\pm0.10$, at $s=20.8$GeV$^2$, $\theta_{\rm CM}=90^\circ$,
for the non-diagonal helicity violating matrix element. If the persistence of
helicity non-conservation is correctly understood as due to independent
scattering processes which do not select mini-hadrons and thus are not
subject to color transparency, helicity conservation should be restored
at the same $Q^2$ in proceeses filtered by  nuclei. One should thus observe
a monotonic decrease of $\rho_{1-1}$ with $A$.

\noindent {\bf \Large Acknowledgements}.

  Most of the work described here was done in
collaboration with John P. Ralston, whom I thank again.

\end{document}